\documentclass[%
 aip,
 amsmath,amssymb,
 reprint,%
]{revtex4-1}

\usepackage{graphicx}
\usepackage{dcolumn}
\usepackage{bm}
\usepackage{physics}
\usepackage[driverfallback=dvipdfm]{hyperref}
\hypersetup{
setpagesize=false,
 bookmarksnumbered=true,%
 bookmarksopen=true,%
 colorlinks=true,%
 linkcolor=blue,
 citecolor=red,}
\usepackage[utf8]{inputenc}
\usepackage[T1]{fontenc}
\usepackage{mathptmx}
\usepackage{color}
\usepackage{etoolbox}

\begin{document}


\title{Nonlinear Seebeck effect in $\rm{Ni_{81}Fe_{19}}|Pt$ at room temperature}
\author{Y. Hirata\textsuperscript{*}}
 \affiliation{Department of Applied Physics, The University of Tokyo, Tokyo 113-8656, Japan}
\author{T. Kikkawa\textsuperscript{**}}%
  \affiliation{Department of Applied Physics, The University of Tokyo, Tokyo 113-8656, Japan}
  \affiliation{Advanced Science Research Center, Japan Atomic Energy Agency, Tokai 319-1195, Japan}
\author{H. Arisawa}%
\affiliation{Department of Applied Physics, The University of Tokyo, Tokyo 113-8656, Japan}
\affiliation{RIKEN Center for Emergent Matter Science (CEMS), Wako 351-0198, Japan}
\author{E. Saitoh}
\affiliation{Department of Applied Physics, The University of Tokyo, Tokyo 113-8656, Japan}
\affiliation{RIKEN Center for Emergent Matter Science (CEMS), Wako 351-0198, Japan}
\affiliation{Institute for AI and Beyond, The University of Tokyo, Tokyo 113-8656, Japan}
\affiliation{WPI Advanced Institute for Materials Research, Tohoku University, Sendai 980-8577, Japan}

\thanks{\textsuperscript{*}Email:1316tbtn@g.ecc.u-tokyo.ac.jp}
\thanks{\textsuperscript{**}Email:kikkawa.takashi@jaea.go.jp}


\begin{abstract}
The nonlinear Seebeck effect, nonlinear conversion of a temperature gradient into an electric current, was observed at room temperature. Based on a second-harmonic lock-in method combined with an a.c. temperature gradient, $\nabla T$, we measured a nonlinear Seebeck voltage in NiFe|Pt bilayers at $300\ \rm K$, the amplitude of which increases in proportion to $(\nabla T)^2$. We also observed that the nonlinear Seebeck voltage increases as the sample length along the $\nabla T$ direction decreases, showing a characteristic scaling law distinct from the conventional linear Seebeck effect. We developed a phenomenological model for the nonlinear Seebeck effect incorporating the spin-current induced modulation of the Seebeck coefficient, which well reproduces the experimental results.
\end{abstract}

\maketitle


In the linear response regime, the current and voltage 
appiled to a sample in a same direction
are proportional to each other, a relationship known as the Ohm's law. On the other hand, nonlinear second-order current can flow in asymmetric systems such as heterojunction, polar or chiral materials\cite{Ideue_natphys,Te_natmat,TokuraNagaosa_natcomm,oshigane_APL}; 
the electron flow can be expanded in a power series with respect to the electric field $E$,
\begin{eqnarray}
{j}_{\rm e}(E)=\sigma_1E+\sigma_2{E}^2+\sigma_3E^3+\cdots,
\label{eq:general_electric_response}
\end{eqnarray}
where $\sigma_i$ is the $i$-th order electric conductivity, and 
the second-order conductivity $\sigma_2$ represents an asymmetric electric response with respect to the direction of $E$. This gives rise to nonreciprocal electric transport,
represented as $|j_{\rm e}(E)|\neq|j_{\rm e}(-E)|$.
Breaking the spatial-inversion and time-reversal symmetry play an important role in the second-order nonlinear transport $\sigma_2$
\cite{Ideue_natphys,Te_natmat,TokuraNagaosa_natcomm,quantummetric_dipole_nat,quantummetric_dipole_nat2}.

\begin{figure}[htb]
\begin{center}
\includegraphics[width=8.5cm]{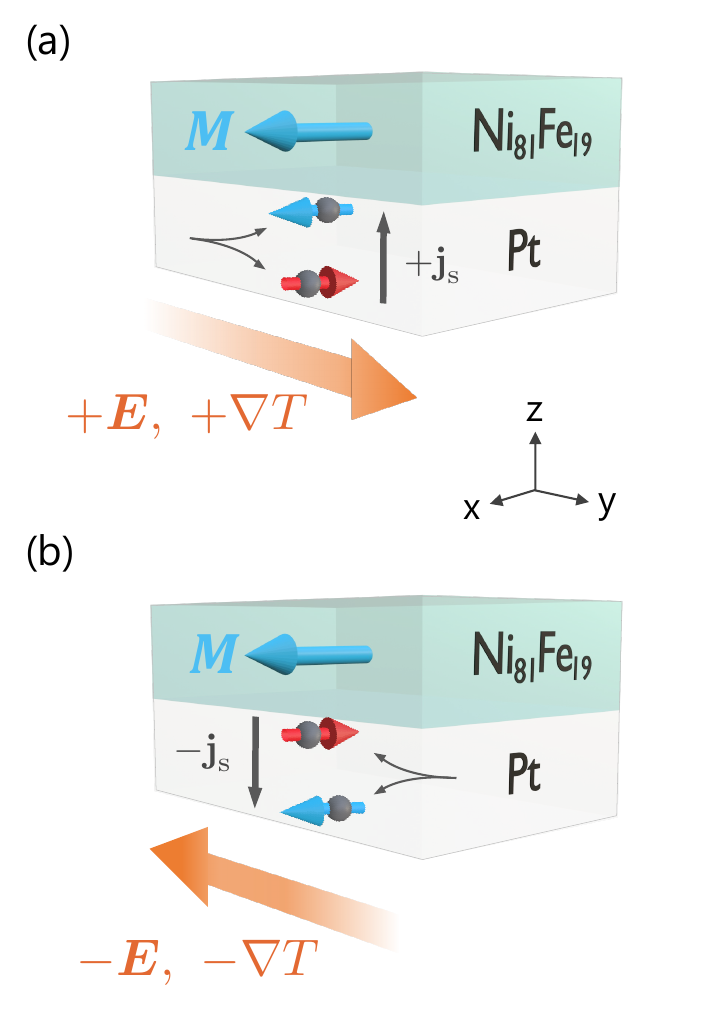}
\end{center}
\caption{Schematic illustrations of the bilayer system consisting of a ferromagnetic metal (FM) NiFe and
a heavy metal (HM) Pt. The red (blue) arrows denote up (down) spins of the electrons.
(a) When a positive electric field $+\bf E\parallel +\hat{\bf y}$ (a temperature gradient $+\nabla T\parallel +\hat{\bf y}$) is applied in the $y$ direction, the spin Hall (Nernst) current $+\bf j_{\rm s}$ is induced in the $z$ direction.
(b) Under a negative electric field $-\bf E\parallel -\hat{\bf y}$ (a temperature gradient $-\nabla T\parallel -\hat{\bf y}$), the direction of the spin Hall (Nernst) current gets reversed to $-\bf j_{\rm s}$. The scattering intensity between the electrons in the HM and magnetization ${\bf M}$ in the FM, therefore, varies depending on the input direction.
}
\label{fig:1}
\end{figure}

Unidirectional spin Hall magnetoresistance (USMR) \cite{USMR_1,USMR_4,USMR_5,SMR_exp,SMR_theory}, one of typical nonlinear transport phenomena, occurs in a bilayer system consisting of a metal and a ferromagnetic layers, which breaks both the spatial-inversion and time-reversal symmetries. When an electric field, ${\bf E}$, is applied to the system, the spin Hall effect (SHE) \cite{spin_Hall} is driven in the metallic layer and induces a nonequilibrium spin accumulation, ${\delta \bm \sigma}_{\rm s}$, at the interface, with the spin polarization parallel to the ${+\hat {\bf x}}$ (${-\hat {\bf x}}$) direction when ${\bf E}\ ||\ {+\hat {\bf y}}$ (${\bf E}\ ||\ {-\hat {\bf y}}$) (see Fig. \ref{fig:1}). 
Then, ${\delta \bm \sigma}_{\rm s}$ interacts with the magnetization ${\bf M}$ in the ferromagnet, which modulates ${\delta \bm \sigma}_{\rm s}$, or the electron spin-dependent chemical potential $\mu_{\rm e}^{\uparrow(\downarrow)}$, affecting the electron conductivity \cite{USMR_1,USMR_2}. Since ${\delta \bm \sigma}_{\rm s}$ depends on the applied ${\bf E}$ direction, this leads to the nonreciprocal or nonlinear conductivity $\sigma_2$, whose sign can be modulated by the ${\bf M}$ direction. USMR has been observed in various junction systems at room temperature\cite{USMR_1,USMR_2}.
 


%
The electric field induced by thermoelectric effects\cite{thermoele_1,thermoele_2,thermoele_3} can be expanded with respect to the temperature gradient $\nabla T$,
\begin{eqnarray}
{E}(\nabla T)=S_1\nabla T+S_2(\nabla T)^2+S_3(\nabla T)^3+\cdots,
\label{eq:general_thermoelectric_response}
\end{eqnarray}
where $S_i$ is the $i$-th order Seebeck coefficient. 
Under the finite $S_2$, nonreciprocal thermopower $|{E}(\nabla T)|\neq |{E}(-\nabla T)|$ can occur.\cite{nakainagaosa_PRB}
Recently, the second-order nonlinear Nernst effect, which generates a thermoelectric voltage in proportional to the square of $\nabla T$, was observed in a bilayer film composed of a type-II superconductor and a ferromagnetic insulator\cite{arisawa_natcomm}.
This nonlinear effect, however, has been limited at an extremely low temperature $(<4\ \rm{K})$ to occur.
For the practical application, it has been eagerly awaited to explore nonlinear thermopower at room temperature.

Here, we report the observation of a nonlinear Seebeck effect at room temperature in a junction system composed of a ferromagnetic metal(FM) NiFe and a heavy metal(HM) Pt, which corresponds to the thermoelectric counterpart of the USMR.


\begin{figure}[htb]
\begin{center}
\includegraphics[width=8.5cm]{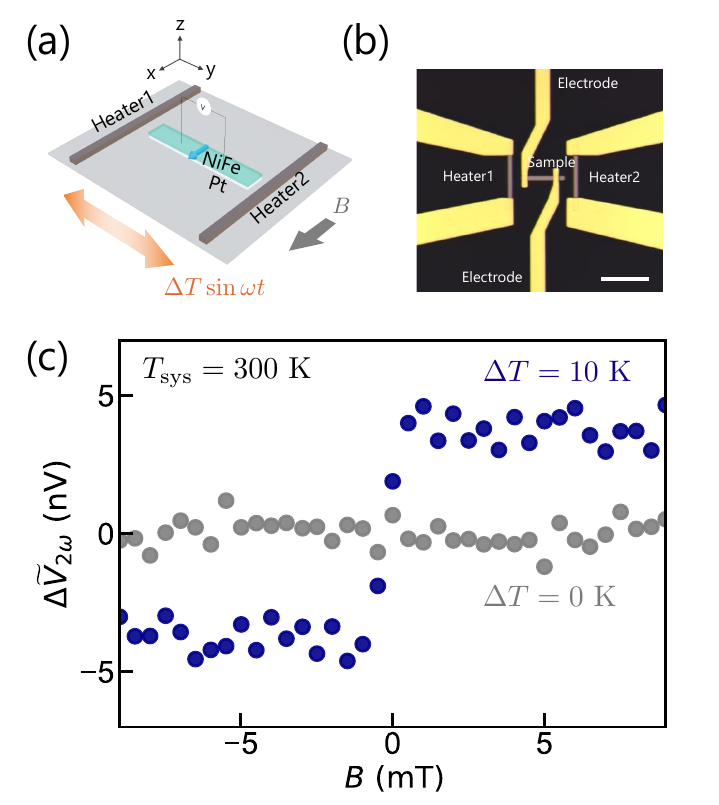}
\end{center}
\caption{
(a) Schematic experimental setup for the nonlinear Seebeck effect. A rectangular-shaped NiFe|Pt bilayer placed between two heaters are formed on an $\rm{Al_2O_3}$ substrate.
(b) A photograph of the sample system used in the experiment taken with an optical microscope. The white bar represents a length of 50 $\rm{\mu m}$.
(c) Second harmonic magneto-Seebeck voltage of the NiFe|Pt bilayer ($l=40~\mu\textrm{m}$) as a function of the magnetic field $B$ along the $x$ axis measured at $\textcolor{black}{T_{\rm sys}} = 300\ \textrm{K}$. 
$\Delta T$ \textcolor{black}{and $T_{\rm sys}$} denote the temperature difference between the sample ends along the $y$ axis \textcolor{black}{and the system temperature, respectively}. The input frequency $\omega/2\pi$ for the a.c. voltage was set to $6.723\ \rm Hz$.
}
\label{fig:2}
\end{figure}

We prepared 
$\rm{Ni}_{81}\rm{Fe}_{19}$ (5 nm)|Pt (5 nm) films, patterned into rectangle-shaped bars with the length of $l=20$, $40$, $60\ \rm{\mu m}$ and the width of $w=5\ \rm{\mu m}$. 
We mainly measured the sample with $l=40\ \rm{\mu m}$, except for sample-length dependence measurements.
Ni$_{81}$Fe$_{19}$ (hereafter referred to as NiFe)  exhibits 
ferromagnetism\cite{Curie_temp},
whereas Pt does high thermo-spin conversion efficiency at room temperature\cite{spinNernst_Pt,spinNernst_W,maekawa_review_spin_current}.
Our experimental set-up is shown in Figs. \ref{fig:2}(a) and \ref{fig:2}(b).
Each of film strips was deposited on ${\rm Al}_2{\rm O}_3$ substrates by magnetron sputtering in a $10^{-1}$ Pa Ar atmosphere and
patterned by photolithography 
and lift-off methods. The deposition rates for the NiFe and Pt layers are $0.014\ \rm nm/s$ and $0.023\ \rm nm/s$, respectively. 
Two heaters (Heater 1 and Heater 2) were made on both sides of the sample on the substrate, each positioned $10\ \rm{\mu m}$ away from it [see Fig. \ref{fig:2}(b)]. Ti/Au electrodes were then made on the sample and heaters for electrical contact.
Upon the application of the temperature gradient $\nabla T$ to the sample along the $y$ direction in Fig. \ref{fig:1}, an interfacial spin accumulation, $\delta\sigma_{\rm s}$, can be thermally induced with the spin polarization parallel to the ${+\hat {\bf x}}\ ({-\hat {\bf x}})$ direction when ${\nabla T}\parallel{+\hat {\bf y}}\ ({\nabla T}\parallel{-\hat {\bf y}})$ (see Fig. \ref{fig:1}) due to the spin Nernst effect\cite{spinNernst_Pt,spinNernst_W,SNE_NE,SNE_theory,hetero_nonlinear} in the Pt layer.
In analogy with the USMR, the Seebeck coefficient can be modulated depending on the magnetization ${\bf M}$ direction, ${\bf M}\parallel{+\hat {\bf x}}$ or ${\bf M}\parallel{-\hat {\bf x}}$, which can be detected as the nonlinear thermoelectric voltage.

To measure the nonlinear Seebeck effect, we employ the second harmonic $(2\omega)$ lock-in detection technique using an input a.c. sinusoidal temperature gradient $\nabla T\propto\sin(\omega t)$. In this method, biased a.c. voltages with the same frequency but different phase are applied to the two heaters as described in Ref. \onlinecite{arisawa_natcomm}.
With the application of the voltage $V_i(t)(i=1,2)$ to the Heater 1 and Heater 2, Joule heating induces a temperature gradient in the in-plane direction as
\begin{eqnarray}
\nabla T(t) \propto r_1V_1(t)^2-r_2V_2(t)^2,
\label{eq:grad-T_original}
\end{eqnarray}
where $r_i$ is a prefactor relating the Joule heating of the Heater $i$ to the temperature gradient.
By applying the biased a.c. voltage $V_i(t)=V_{i,0}+V_{i, \rm amp}\sin(\omega t+\phi_i)$ with
$\phi_1=0$ and $\phi_2=\pi$, the input temperature gradient oscillates at the frequency of $\omega/2\pi$:
\begin{eqnarray}
\nabla T(t) \propto (r_1V_{1,0}V_{1,\rm amp}+r_2V_{2,0}V_{2,\rm amp})\sin\omega t+Const.
\label{eq:grad-T}
\end{eqnarray}
For details on the optimization of the  $V_{i,0}$ and $V_{i, \rm amp}$ values and on the generation of the sinusoidal a.c. temperature gradient, see Sec. 1 in Supplementary Information. The typical voltage parameters applied to Heater $i$ are as follows: the d.c. component $V_{i, \rm dc}$ of $0.1-5$ V, the root-mean-square (rms) amplitude $V_{i, \rm amp}$ of $0.1-3.5$ V, and the frequency of 6.723 Hz.
\textcolor{black}{We note that two synchronized heat sources are required to create a sinusoidal a.c. temperature gradient, since the direction of Joule heat current cannot be reversed with a single source, unlike an a.c. electric current used to detect nonlinear electric transport\cite{Ideue_natphys}.}
\textcolor{black}{
All the measurements were performed under vacuum, with the temperature of the system (the sample stage) maintained at $T_{\rm sys} = 300\ \textrm{K}$ using a PPMS (Quantum Design Inc.).
}

\begin{figure}[htb]
\begin{center}
\includegraphics[width=8.5cm]{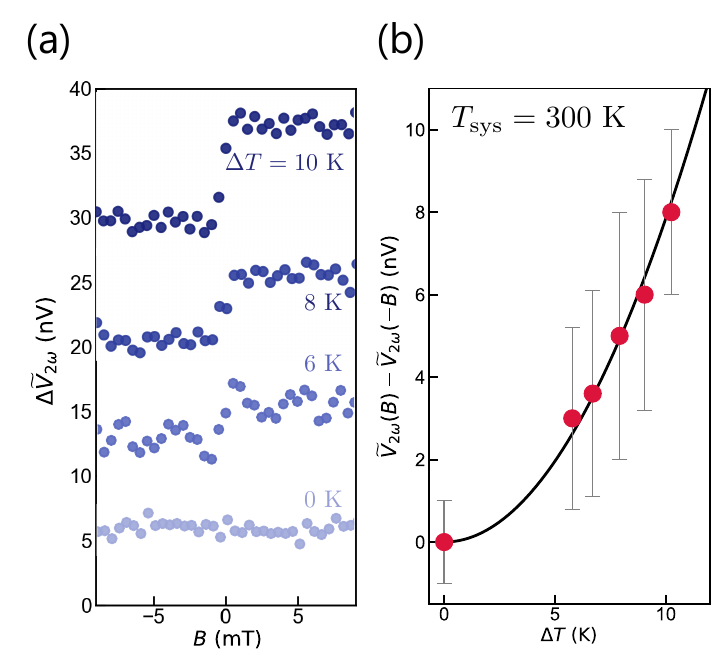}
\end{center}
\caption{
(a) Second harmonic magneto-Seebeck voltage of the NiFe|Pt bilayer  ($l=40~\mu\textrm{m}$) at the input temperature differences of $\Delta T = 
0$, $6$, $8$, and $10\ \rm K$.
The offsets are added to each of the $\widetilde{V}_{2\omega}-B$ scans for clarity.
(b) $\Delta T$ dependence of the second harmonic voltage difference $\widetilde{V}_{2\omega}$ between $+B(=5\ \rm{mT})$ and $-B(=-5\ \rm{mT})$.
\textcolor{black}{Gray bars denote error bars of each measurement.}
}
\label{fig:3}
\end{figure}

We measured the second-harmonic Seebeck (longitudinal) voltage $V_{y,2\omega}$ via the lock-in measurement\cite{Ideue_natphys,lockin_1,lockin_3,lockin_4}. 
Figure \ref{fig:2}(c) shows the second harmonic voltage as a function of the magnetic field $B$ along the $x$ direction for the NiFe|Pt sample at 300 K. The sign of the signal changes depending on the $B$ direction, which corresponds to the magnetization curve in the NiFe layer. The result is consistent with the USMR, where the scattering intensity of electrons in the Pt carrying spin $\vb s$ by the magnetization $\bf M$ in the NiFe depends on the angle between $\vb M$ and $\vb s$. The vertical axis in Fig. \ref{fig:2}(c) represents $\widetilde{V}_{2\omega}:=V_{y,2\omega}(V_{1,0},V_{1,\rm amp},V_{2,0},V_{2,\rm amp})-V_{y,2\omega}(0,V_{1,\rm amp},0,V_{2,\rm amp})$, by which a possible residue of the $2\omega$-$\nabla T$ component can be eliminated (see Sec. 2 in Supplementary Information).
$\Delta$ means that the offset component is subtracted. All the $V-B$ data are anti-symmetrized with respect to the magnetic field $B$.


Figure \ref{fig:3}(a) shows the $B$ dependence of the second harmonic voltage $\widetilde{V}_{2\omega}$ at several $\Delta T$ values.
The signal amplitude $\widetilde{V}_{2\omega}$ (the step height of $V_{2\omega}$ at $B=0$) monotonically increases with the applied temperature difference  $\Delta T$ between the sample edges.
As shown in Fig. \ref{fig:3}(b), the $\widetilde{V}_{2\omega}$ is proportional to the input temperature-difference squared $(\Delta T)^2$ in the NiFe$|$Pt bilayer, \textcolor{black}{even when considering the error bars}.
$V_{2\omega}$ was found to exhibit a characteristic response to the relative phase ($\phi$) between the applied a.c. voltage, $V_{2\omega}\propto \sin^2\qty(\phi/2)\cos\phi$, consistent with the scenario of the nonlinear thermoelectric effect  \cite{arisawa_natcomm}. The signs of the nonlinear Seebeck coefficient were found to be opposite between the NiFe|Pt and Pt|NiFe samples, where the stacking order is reversed, showing that the signal originates from the breaking of inversion symmetry of the sample. For details, see Secs. 3 and 4 in Supplementary information.

\begin{figure}[htb]
\begin{center}
\includegraphics[width=8.5cm]{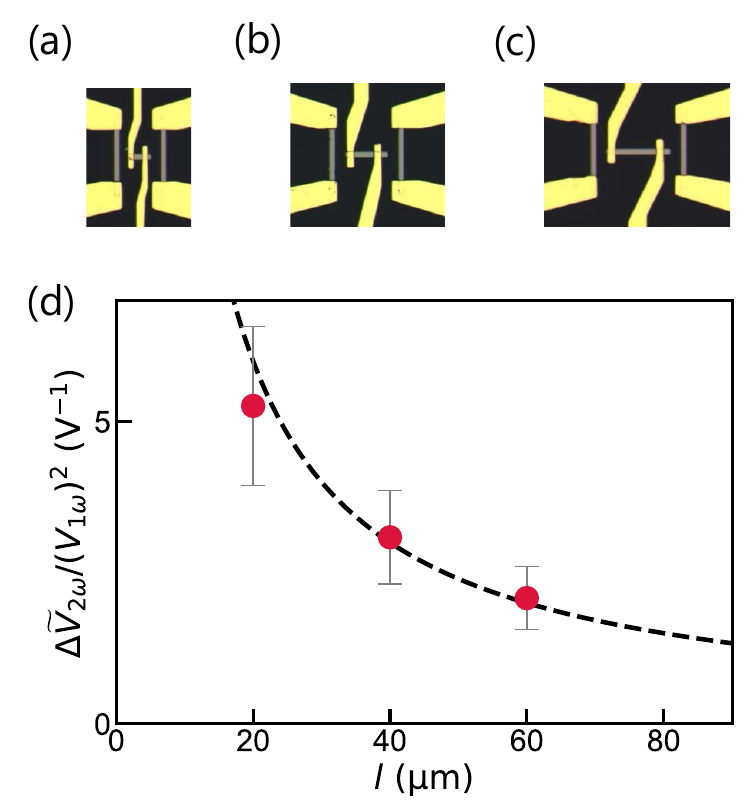}
\end{center}
\caption{
(a-c) Photographs of the measurement system taken with an optical microscope, where the sample length  $l$ is (a) $20$, (b) $40$, and (c) $60\ \rm \mu m$, respectively.
(d) Sample-length dependence of the nonlinear Seebeck voltage normalized by the square of the linear Seebeck voltage. The dashed line represents a fitting curve of the form 
$\propto l^{-1}$. \textcolor{black}{Gray bars denote error bars of each measurement.}
All the plots are measured at room temperature $\textcolor{black}{T_{\rm sys}}=300\ \rm K$.
}
\label{fig:4}
\end{figure}

We prepared samples of various sample lengths $l$ in a single deposition and compared 
their thermoelectric response. The sample photographs are shown in Figs. \ref{fig:4}(a-c). 
The ratio of the measured $2\omega$ voltage to the $1\omega$ voltage squared $\widetilde{V}_{2\omega}/(V_{1\omega})^2$ decreases with increasing $l$.
This result is consistent with the $l$ dependence of the nonlinear Seebeck signal; by multiplying $l$ to Eq. (\ref{eq:general_thermoelectric_response}), 
\begin{eqnarray}
    \Delta V(\nabla T)=S_1\Delta T+S_2\frac{(\Delta T)^2}{l}+S_3\frac{(\Delta T)^3}{l^2}+\cdots,
    \label{lengthdep_thermoelectric}
\end{eqnarray}
where the first term (linear thermopower) $V_{1\omega}:= S_1\Delta T$ is independent on $l$. On the other hand, the second term (nonlinear thermopower) $V_{2\omega}:=S_2(\Delta T)^2/l$ is proportional to $l^{-1}$. Therefore, $V_{2\omega}/(V_{1\omega})^2$ should be proportional to $l^{-1}$, which well reproduces the experimental data [see Fig. \ref{fig:4}(d)].
This scaling law for the nonlinear thermoelectric voltage may be applied to improving the efficiency of the effect.

We formulated a phenomenological model to describe the observed nonlinear Seebeck effect in a FM|HM bilayer, based on a USMR model\cite{USMR_1,USMR_2,USMR_3,Steven_USMR_theory}.
Under the temperature gradient $\nabla T$, the spin Nernst effect\cite{spinNernst_Pt,spinNernst_W} in the HM layer induces an interfacial spin accumulation, $\delta\sigma_{\rm s}$, depending on the direction of $\nabla T$.
\begin{figure}[htb]
\begin{center}
\includegraphics[width=8.5cm]{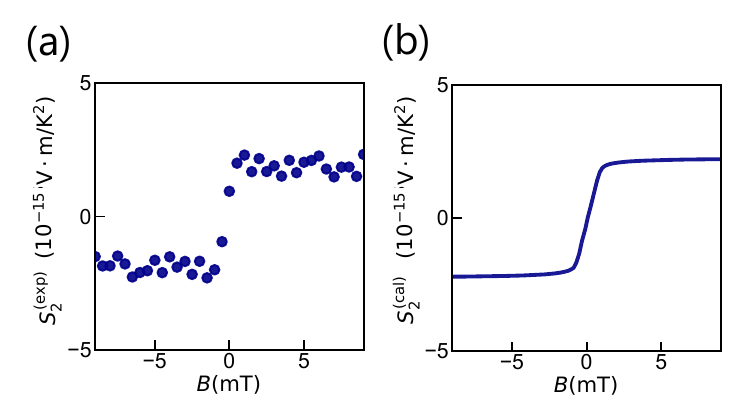}
\end{center}
\caption{
Comparison between the $B$ dependence of (a) the experimental $S_2(B)\approx\widetilde{V}_{2\omega}l/(\Delta T)^2$ for the NiFe|Pt bilayer  ($l=40~\mu\textrm{m}$)  and (b) the calculated $S_2$ based on Eq. (\ref{eq:S_2}). 
}
\label{fig:5}
\end{figure}
Subsequently, the spin diffusion into the FM layer modulates the electron chemical potential $\Delta\mu_{\rm e}^{\uparrow(\downarrow)}(\propto \nabla T)$ in the FM layer. Because the linear Seebeck coefficient $S_{1,\rm F}^{\uparrow(\downarrow)}$ and electric conductivity $\sigma_{1,\rm F}^{\uparrow(\downarrow)}$ depend on the chemical potential $\mu_{\rm e}$ in the FM layer, the thermoelectric response can be modulated in the second-order nonlinear regime.
The expression of this nonlinear coefficient $S_2$ reads
\begin{eqnarray}
S_2\partial_y T\approx \frac{d_{\rm F}\Delta\mu_{e}}{\sigma_{1,\rm H}d_{\rm H}+\sigma_{1,\rm F}d_{\rm F}}\left.\frac{\partial}{\partial E}(\sigma_{1,\rm F}^{\uparrow}S_{1,\rm F}^{\uparrow}+\sigma_{1,\rm F}^{\downarrow}S_{1,\rm F}^{\downarrow})\right|_{E=\mu_e},
\label{eq:S_2}
\end{eqnarray}
\begin{eqnarray}
    \Delta\mu_{\rm e}=
\frac{P_s\theta_{\rm SN}S_{1,\rm H}\nabla T\left(\cosh \frac{d_{\rm H}}{\lambda_{\rm H}}-1\right)}{(1-P_s^2)\frac{\sigma_{1,\rm F}}{\lambda_{\rm F}\sigma_{1,\rm H}}+\frac{1}{\lambda_{\rm H}}\sinh \frac{d_{\rm H}}{\lambda_{\rm H}}},
\label{eq:Delta_mu}
\end{eqnarray}
where $\theta_{\rm SN},\ d_{\rm F(H)},\ \lambda_{\rm F(H)},\ P_s(:=(D^{\uparrow}-D^{\downarrow})/(D^{\uparrow}+D^{\downarrow}))$, $D^{\uparrow(\downarrow)}$, $\sigma_{\rm 1, H}$ and $S_{1,\rm H}$ denote the spin-Nernst angle in the HM layer, spin diffusion length and thickness of the FM (HM) layer, spin polarizability and spin-dependent state density of the FM layer, linear electric conductivity and Seebeck coefficient of the HM layer, respectively.
Equation (\ref{eq:Delta_mu}) is a thermoelectric extension of the USMR, which we call the unidirectional spin Nernst magneto-thermopower (USMT). The model well reproduces the experimental result [see Fig. \ref{fig:5}(a) and \ref{fig:5}(b)] \textcolor{black}{and suggests that $S_2$ values may be enhanced with a larger spin Nernst angle $\theta_{\rm SN}$ and a higher linear thermopower $S_{\rm 1,H}$. Suitable candidates for enhancing $S_2$ include certain topological insulators\cite{Bi2Se3_Seebeck,Bi2Se3_spinNernst,Bi88Se12_spinNernst}.
In addition to material substitution, alternative device architectures may also be effective. For instance, low-symmetry bulk crystals or lateral heterostructures have recently shown strong nonlinear electric conductivity $\sigma_2$ even at room temperature\cite{strong_nonlinear_Hall,strong_diode_electrical_and_thermal}, suggesting that large nonlinear thermoelectric coefficients $S_2$ may be achievable in such systems.
We note that materials with a large $S_2$ can directly generate a d.c. voltage from temperature fluctuations, without requiring precisely controlled sinusoidal temperature gradients for measurement.
}

In summary, 
we observed the nonlinear Seebeck effect in $\rm{Ni}_{81}\rm{Fe}_{19}$|Pt films at room temperature based on the second harmonic lock-in method.
The nonlinear voltage is quadratic with respect to the input temperature gradient and inversely proportional to the sample length.
The experimental result can be reproduced by a calculation, in which the modulation of the electron chemical potential by the spin Nernst effect is taken into consideration.
The effect can be applied to making an electric power generator or a sensor for detecting temperature fluctuations at room temperature 
\textcolor{black}{that are abundant in our surroundings yet have so far eluded observation.}

See the supplementary material for the practical method to generate a sinusoidal a.c. temperature gradient; the method for eliminating a possible linear-thermoelectric contribution to the $2\omega$ lock-in voltage; consideration on the exclusive lock-in detection for the nonlinear thermoelectric voltage; and the effect of stacking order on the relative phase $\phi$ dependence of the $2\omega$ lock-in voltage.

The authors thank Y. Fujimoto and Z. Chen for their help on sample fabrication and S. Daimon for fruitful discussions about the experimental results. This work was supported by CREST (Nos. JPMJCR20C1 and JPMJCR20T2) from JST, Japan, Grant-in-Aid for Scientific Research (S) (No. JP19H05600), Grant-in-Aid for Scientific Research (B) (No. JP24K01326), Grant-in-Aid for Transformative Research Areas (No. JP22H05114) from JSPS KAKENHI, Japan, and Sumitomo Chemical.

Y.H., T.K., and E.S constructed the experimental setup; Y.H. carried out the device fabrication and patterning, performed the experiments, analyzed the data and formulated the theoretical model.
Y.H. and T.K. wrote the paper with review and input from H.A. and E.S.
T.K. and E.S. conceived the study. E.S. supervised the project.
All authors contributed to the discussion of the data in the manuscript and Supplementary Information.

The data measured in this study are available from the corresponding author upon reasonable request.

\bibliography{reference}

\begin{thebibliography}{35}%
\makeatletter
\providecommand \@ifxundefined [1]{%
 \@ifx{#1\undefined}
}%
\providecommand \@ifnum [1]{%
 \ifnum #1\expandafter \@firstoftwo
 \else \expandafter \@secondoftwo
 \fi
}%
\providecommand \@ifx [1]{%
 \ifx #1\expandafter \@firstoftwo
 \else \expandafter \@secondoftwo
 \fi
}%
\providecommand \natexlab [1]{#1}%
\providecommand \enquote  [1]{``#1''}%
\providecommand \bibnamefont  [1]{#1}%
\providecommand \bibfnamefont [1]{#1}%
\providecommand \citenamefont [1]{#1}%
\providecommand \href@noop [0]{\@secondoftwo}%
\providecommand \href [0]{\begingroup \@sanitize@url \@href}%
\providecommand \@href[1]{\@@startlink{#1}\@@href}%
\providecommand \@@href[1]{\endgroup#1\@@endlink}%
\providecommand \@sanitize@url [0]{\catcode `\\12\catcode `\$12\catcode `\&12\catcode `\#12\catcode `\^12\catcode `\_12\catcode `\%12\relax}%
\providecommand \@@startlink[1]{}%
\providecommand \@@endlink[0]{}%
\providecommand \url  [0]{\begingroup\@sanitize@url \@url }%
\providecommand \@url [1]{\endgroup\@href {#1}{\urlprefix }}%
\providecommand \urlprefix  [0]{URL }%
\providecommand \Eprint [0]{\href }%
\providecommand \doibase [0]{http://dx.doi.org/}%
\providecommand \selectlanguage [0]{\@gobble}%
\providecommand \bibinfo  [0]{\@secondoftwo}%
\providecommand \bibfield  [0]{\@secondoftwo}%
\providecommand \translation [1]{[#1]}%
\providecommand \BibitemOpen [0]{}%
\providecommand \bibitemStop [0]{}%
\providecommand \bibitemNoStop [0]{.\EOS\space}%
\providecommand \EOS [0]{\spacefactor3000\relax}%
\providecommand \BibitemShut  [1]{\csname bibitem#1\endcsname}%
\let\auto@bib@innerbib\@empty
\bibitem [{\citenamefont {Ideue}\ \emph {et~al.}(2017)\citenamefont {Ideue}, \citenamefont {Hamamoto}, \citenamefont {Koshikawa}, \citenamefont {Ezawa}, \citenamefont {Shimizu}, \citenamefont {Kaneko}, \citenamefont {Tokura}, \citenamefont {Nagaosa},\ and\ \citenamefont {Iwasa}}]{Ideue_natphys}%
  \BibitemOpen
  \bibfield  {author} {\bibinfo {author} {\bibfnamefont {T.}~\bibnamefont {Ideue}}, \bibinfo {author} {\bibfnamefont {K.}~\bibnamefont {Hamamoto}}, \bibinfo {author} {\bibfnamefont {S.}~\bibnamefont {Koshikawa}}, \bibinfo {author} {\bibfnamefont {M.}~\bibnamefont {Ezawa}}, \bibinfo {author} {\bibfnamefont {S.}~\bibnamefont {Shimizu}}, \bibinfo {author} {\bibfnamefont {Y.}~\bibnamefont {Kaneko}}, \bibinfo {author} {\bibfnamefont {Y.}~\bibnamefont {Tokura}}, \bibinfo {author} {\bibfnamefont {N.}~\bibnamefont {Nagaosa}}, \ and\ \bibinfo {author} {\bibfnamefont {Y.}~\bibnamefont {Iwasa}},\ }\bibfield  {title} {\enquote {\bibinfo {title} {Bulk rectification effect in a polar semiconductor},}\ }\href {\doibase https://doi.org/10.1038/nphys4056} {\bibfield  {journal} {\bibinfo  {journal} {Nat. Phys.}\ }\textbf {\bibinfo {volume} {13}},\ \bibinfo {pages} {578--583} (\bibinfo {year} {2017})}\BibitemShut {NoStop}%
\bibitem [{\citenamefont {Calavalle}\ \emph {et~al.}(2022)\citenamefont {Calavalle}, \citenamefont {Suárez-Rodríguez}, \citenamefont {Martín-García}, \citenamefont {Johansson}, \citenamefont {Vaz}, \citenamefont {Yang}, \citenamefont {Maznichenko}, \citenamefont {Ostanin}, \citenamefont {Mateo-Alonso}, \citenamefont {Chuvilin}, \citenamefont {Mertig}, \citenamefont {Gobbi}, \citenamefont {Casanova},\ and\ \citenamefont {Hueso}}]{Te_natmat}%
  \BibitemOpen
  \bibfield  {author} {\bibinfo {author} {\bibfnamefont {F.}~\bibnamefont {Calavalle}}, \bibinfo {author} {\bibfnamefont {M.}~\bibnamefont {Suárez-Rodríguez}}, \bibinfo {author} {\bibfnamefont {B.}~\bibnamefont {Martín-García}}, \bibinfo {author} {\bibfnamefont {A.}~\bibnamefont {Johansson}}, \bibinfo {author} {\bibfnamefont {D.~C.}\ \bibnamefont {Vaz}}, \bibinfo {author} {\bibfnamefont {H.}~\bibnamefont {Yang}}, \bibinfo {author} {\bibfnamefont {I.~V.}\ \bibnamefont {Maznichenko}}, \bibinfo {author} {\bibfnamefont {S.}~\bibnamefont {Ostanin}}, \bibinfo {author} {\bibfnamefont {A.}~\bibnamefont {Mateo-Alonso}}, \bibinfo {author} {\bibfnamefont {A.}~\bibnamefont {Chuvilin}}, \bibinfo {author} {\bibfnamefont {I.}~\bibnamefont {Mertig}}, \bibinfo {author} {\bibfnamefont {M.}~\bibnamefont {Gobbi}}, \bibinfo {author} {\bibfnamefont {F.}~\bibnamefont {Casanova}}, \ and\ \bibinfo {author} {\bibfnamefont {L.~E.}\ \bibnamefont {Hueso}},\ }\bibfield  {title} {\enquote {\bibinfo {title} {Gate-tuneable and
  chirality-dependent charge-to-spin conversion in tellurium nanowires},}\ }\href {\doibase https://doi.org/10.1038/s41563-022-01211-7} {\bibfield  {journal} {\bibinfo  {journal} {Nat. Mater.}\ }\textbf {\bibinfo {volume} {21}},\ \bibinfo {pages} {526--532} (\bibinfo {year} {2022})}\BibitemShut {NoStop}%
\bibitem [{\citenamefont {Tokura}\ and\ \citenamefont {Nagaosa}(2018)}]{TokuraNagaosa_natcomm}%
  \BibitemOpen
  \bibfield  {author} {\bibinfo {author} {\bibfnamefont {Y.}~\bibnamefont {Tokura}}\ and\ \bibinfo {author} {\bibfnamefont {N.}~\bibnamefont {Nagaosa}},\ }\bibfield  {title} {\enquote {\bibinfo {title} {Nonreciprocal responses from non-centrosymmetric quantum materials},}\ }\href {\doibase https://doi.org/10.1038/s41467-018-05759-4} {\bibfield  {journal} {\bibinfo  {journal} {Nat. Commun.}\ }\textbf {\bibinfo {volume} {9}},\ \bibinfo {pages} {3740} (\bibinfo {year} {2018})}\BibitemShut {NoStop}%
\bibitem [{\citenamefont {Oshigane}, \citenamefont {Arisawa},\ and\ \citenamefont {Saitoh}(2025)}]{oshigane_APL}%
  \BibitemOpen
  \bibfield  {author} {\bibinfo {author} {\bibfnamefont {K.}~\bibnamefont {Oshigane}}, \bibinfo {author} {\bibfnamefont {H.}~\bibnamefont {Arisawa}}, \ and\ \bibinfo {author} {\bibfnamefont {E.}~\bibnamefont {Saitoh}},\ }\bibfield  {title} {\enquote {\bibinfo {title} {Nonreciprocal and nonlinear transport and spontaneous voltage generation in $\rm{MoGe/Ni_{81}Fe_{19}}$},}\ }\href {\doibase https://doi.org/10.1063/5.0237520} {\bibfield  {journal} {\bibinfo  {journal} {Appl. Phys. Lett.}\ }\textbf {\bibinfo {volume} {126}},\ \bibinfo {pages} {022402} (\bibinfo {year} {2025})}\BibitemShut {NoStop}%
\bibitem [{\citenamefont {Wang}\ \emph {et~al.}(2023)\citenamefont {Wang}, \citenamefont {Kaplan}, \citenamefont {Zhang}, \citenamefont {Holder}, \citenamefont {Cao}, \citenamefont {Wang}, \citenamefont {Zhou}, \citenamefont {Zhou}, \citenamefont {Jiang}, \citenamefont {Zhang}, \citenamefont {Ru}, \citenamefont {Cai}, \citenamefont {Watanabe}, \citenamefont {Taniguchi}, \citenamefont {Yan},\ and\ \citenamefont {Gao}}]{quantummetric_dipole_nat}%
  \BibitemOpen
  \bibfield  {author} {\bibinfo {author} {\bibfnamefont {N.}~\bibnamefont {Wang}}, \bibinfo {author} {\bibfnamefont {D.}~\bibnamefont {Kaplan}}, \bibinfo {author} {\bibfnamefont {Z.}~\bibnamefont {Zhang}}, \bibinfo {author} {\bibfnamefont {T.}~\bibnamefont {Holder}}, \bibinfo {author} {\bibfnamefont {N.}~\bibnamefont {Cao}}, \bibinfo {author} {\bibfnamefont {A.}~\bibnamefont {Wang}}, \bibinfo {author} {\bibfnamefont {X.}~\bibnamefont {Zhou}}, \bibinfo {author} {\bibfnamefont {F.}~\bibnamefont {Zhou}}, \bibinfo {author} {\bibfnamefont {Z.}~\bibnamefont {Jiang}}, \bibinfo {author} {\bibfnamefont {C.}~\bibnamefont {Zhang}}, \bibinfo {author} {\bibfnamefont {S.}~\bibnamefont {Ru}}, \bibinfo {author} {\bibfnamefont {H.}~\bibnamefont {Cai}}, \bibinfo {author} {\bibfnamefont {K.}~\bibnamefont {Watanabe}}, \bibinfo {author} {\bibfnamefont {T.}~\bibnamefont {Taniguchi}}, \bibinfo {author} {\bibfnamefont {B.}~\bibnamefont {Yan}}, \ and\ \bibinfo {author} {\bibfnamefont {W.}~\bibnamefont {Gao}},\ }\bibfield  {title}
  {\enquote {\bibinfo {title} {Quantum-metric-induced nonlinear transport in a topological antiferromagnet},}\ }\href {\doibase https://doi.org/10.1038/s41586-023-06363-3} {\bibfield  {journal} {\bibinfo  {journal} {Nature}\ }\textbf {\bibinfo {volume} {621}},\ \bibinfo {pages} {487--492} (\bibinfo {year} {2023})}\BibitemShut {NoStop}%
\bibitem [{\citenamefont {Li}\ \emph {et~al.}(2024)\citenamefont {Li}, \citenamefont {Zhang}, \citenamefont {Zhou}, \citenamefont {Ma}, \citenamefont {Lei}, \citenamefont {Jin}, \citenamefont {He}, \citenamefont {Li}, \citenamefont {Law},\ and\ \citenamefont {Wang}}]{quantummetric_dipole_nat2}%
  \BibitemOpen
  \bibfield  {author} {\bibinfo {author} {\bibfnamefont {H.}~\bibnamefont {Li}}, \bibinfo {author} {\bibfnamefont {C.}~\bibnamefont {Zhang}}, \bibinfo {author} {\bibfnamefont {C.}~\bibnamefont {Zhou}}, \bibinfo {author} {\bibfnamefont {C.}~\bibnamefont {Ma}}, \bibinfo {author} {\bibfnamefont {X.}~\bibnamefont {Lei}}, \bibinfo {author} {\bibfnamefont {Z.}~\bibnamefont {Jin}}, \bibinfo {author} {\bibfnamefont {H.}~\bibnamefont {He}}, \bibinfo {author} {\bibfnamefont {B.}~\bibnamefont {Li}}, \bibinfo {author} {\bibfnamefont {K.~T.}\ \bibnamefont {Law}}, \ and\ \bibinfo {author} {\bibfnamefont {J.}~\bibnamefont {Wang}},\ }\bibfield  {title} {\enquote {\bibinfo {title} {Quantum geometry quadrupole-induced third-order nonlinear transport in antiferromagnetic topological insulator $\rm{MnBi}_2\rm{Te}_4$},}\ }\href {\doibase https://doi.org/10.1038/s41467-024-52206-8} {\bibfield  {journal} {\bibinfo  {journal} {Nat. Commun.}\ }\textbf {\bibinfo {volume} {15}},\ \bibinfo {pages} {7779} (\bibinfo {year}
  {2024})}\BibitemShut {NoStop}%
\bibitem [{\citenamefont {Avci}\ \emph {et~al.}(2015)\citenamefont {Avci}, \citenamefont {Garello}, \citenamefont {Ghosh}, \citenamefont {Gabureac}, \citenamefont {Alvarado},\ and\ \citenamefont {Gambardella}}]{USMR_1}%
  \BibitemOpen
  \bibfield  {author} {\bibinfo {author} {\bibfnamefont {C.~O.}\ \bibnamefont {Avci}}, \bibinfo {author} {\bibfnamefont {K.}~\bibnamefont {Garello}}, \bibinfo {author} {\bibfnamefont {A.}~\bibnamefont {Ghosh}}, \bibinfo {author} {\bibfnamefont {M.}~\bibnamefont {Gabureac}}, \bibinfo {author} {\bibfnamefont {S.~F.}\ \bibnamefont {Alvarado}}, \ and\ \bibinfo {author} {\bibfnamefont {P.}~\bibnamefont {Gambardella}},\ }\bibfield  {title} {\enquote {\bibinfo {title} {Unidirectional spin {H}all magnetoresistance in ferromagnet/ normal metal bilayers},}\ }\href {\doibase https://doi.org/10.1038/nphys3356} {\bibfield  {journal} {\bibinfo  {journal} {Nat. Phys.}\ }\textbf {\bibinfo {volume} {11}},\ \bibinfo {pages} {570--575} (\bibinfo {year} {2015})}\BibitemShut {NoStop}%
\bibitem [{\citenamefont {Kim}\ \emph {et~al.}(2019)\citenamefont {Kim}, \citenamefont {Li}, \citenamefont {Kim}, \citenamefont {Moriyama}, \citenamefont {Koyama}, \citenamefont {Chiba}, \citenamefont {Lee}, \citenamefont {Lee},\ and\ \citenamefont {Ono}}]{USMR_4}%
  \BibitemOpen
  \bibfield  {author} {\bibinfo {author} {\bibfnamefont {K.~J.}\ \bibnamefont {Kim}}, \bibinfo {author} {\bibfnamefont {T.}~\bibnamefont {Li}}, \bibinfo {author} {\bibfnamefont {S.}~\bibnamefont {Kim}}, \bibinfo {author} {\bibfnamefont {T.}~\bibnamefont {Moriyama}}, \bibinfo {author} {\bibfnamefont {T.}~\bibnamefont {Koyama}}, \bibinfo {author} {\bibfnamefont {D.}~\bibnamefont {Chiba}}, \bibinfo {author} {\bibfnamefont {K.~J.}\ \bibnamefont {Lee}}, \bibinfo {author} {\bibfnamefont {H.~W.}\ \bibnamefont {Lee}}, \ and\ \bibinfo {author} {\bibfnamefont {T.}~\bibnamefont {Ono}},\ }\bibfield  {title} {\enquote {\bibinfo {title} {Possible contribution of high-energy magnons to unidirectional magnetoresistance in metallic bilayers},}\ }\href {\doibase https://doi.org/10.7567/1882-0786/ab1b54} {\bibfield  {journal} {\bibinfo  {journal} {Appl. Phys. Express}\ }\textbf {\bibinfo {volume} {12}},\ \bibinfo {pages} {063001} (\bibinfo {year} {2019})}\BibitemShut {NoStop}%
\bibitem [{\citenamefont {Yasuda}\ \emph {et~al.}(2016)\citenamefont {Yasuda}, \citenamefont {Tsukazaki}, \citenamefont {Yoshimi}, \citenamefont {Takahashi}, \citenamefont {Kawasaki},\ and\ \citenamefont {Tokura}}]{USMR_5}%
  \BibitemOpen
  \bibfield  {author} {\bibinfo {author} {\bibfnamefont {K.}~\bibnamefont {Yasuda}}, \bibinfo {author} {\bibfnamefont {A.}~\bibnamefont {Tsukazaki}}, \bibinfo {author} {\bibfnamefont {R.}~\bibnamefont {Yoshimi}}, \bibinfo {author} {\bibfnamefont {K.~S.}\ \bibnamefont {Takahashi}}, \bibinfo {author} {\bibfnamefont {M.}~\bibnamefont {Kawasaki}}, \ and\ \bibinfo {author} {\bibfnamefont {Y.}~\bibnamefont {Tokura}},\ }\bibfield  {title} {\enquote {\bibinfo {title} {Large unidirectional magnetoresistance in a magnetic topological insulator},}\ }\href {\doibase 10.1103/PhysRevLett.117.127202} {\bibfield  {journal} {\bibinfo  {journal} {Phys. Rev. Lett.}\ }\textbf {\bibinfo {volume} {117}},\ \bibinfo {pages} {127202} (\bibinfo {year} {2016})}\BibitemShut {NoStop}%
\bibitem [{\citenamefont {Kim}\ \emph {et~al.}(2016)\citenamefont {Kim}, \citenamefont {Sheng}, \citenamefont {Takahashi}, \citenamefont {Mitani},\ and\ \citenamefont {Hayashi}}]{SMR_exp}%
  \BibitemOpen
  \bibfield  {author} {\bibinfo {author} {\bibfnamefont {J.}~\bibnamefont {Kim}}, \bibinfo {author} {\bibfnamefont {P.}~\bibnamefont {Sheng}}, \bibinfo {author} {\bibfnamefont {S.}~\bibnamefont {Takahashi}}, \bibinfo {author} {\bibfnamefont {S.}~\bibnamefont {Mitani}}, \ and\ \bibinfo {author} {\bibfnamefont {M.}~\bibnamefont {Hayashi}},\ }\bibfield  {title} {\enquote {\bibinfo {title} {Spin {H}all magnetoresistance in metallic bilayers},}\ }\href {\doibase 10.1103/PhysRevLett.116.097201} {\bibfield  {journal} {\bibinfo  {journal} {Phys. Rev. Lett.}\ }\textbf {\bibinfo {volume} {116}},\ \bibinfo {pages} {097201} (\bibinfo {year} {2016})}\BibitemShut {NoStop}%
\bibitem [{\citenamefont {Chen}\ \emph {et~al.}(2013)\citenamefont {Chen}, \citenamefont {Takahashi}, \citenamefont {Nakayama}, \citenamefont {Althammer}, \citenamefont {Goennenwein}, \citenamefont {Saitoh},\ and\ \citenamefont {Bauer}}]{SMR_theory}%
  \BibitemOpen
  \bibfield  {author} {\bibinfo {author} {\bibfnamefont {Y.-T.}\ \bibnamefont {Chen}}, \bibinfo {author} {\bibfnamefont {S.}~\bibnamefont {Takahashi}}, \bibinfo {author} {\bibfnamefont {H.}~\bibnamefont {Nakayama}}, \bibinfo {author} {\bibfnamefont {M.}~\bibnamefont {Althammer}}, \bibinfo {author} {\bibfnamefont {S.~T.~B.}\ \bibnamefont {Goennenwein}}, \bibinfo {author} {\bibfnamefont {E.}~\bibnamefont {Saitoh}}, \ and\ \bibinfo {author} {\bibfnamefont {G.~E.~W.}\ \bibnamefont {Bauer}},\ }\bibfield  {title} {\enquote {\bibinfo {title} {Theory of spin hall magnetoresistance},}\ }\href {\doibase 10.1103/PhysRevB.87.144411} {\bibfield  {journal} {\bibinfo  {journal} {Phys. Rev. B}\ }\textbf {\bibinfo {volume} {87}},\ \bibinfo {pages} {144411} (\bibinfo {year} {2013})}\BibitemShut {NoStop}%
\bibitem [{\citenamefont {Sinova}\ \emph {et~al.}(2015)\citenamefont {Sinova}, \citenamefont {Valenzuela}, \citenamefont {Wunderlich}, \citenamefont {Back},\ and\ \citenamefont {Jungwirth}}]{spin_Hall}%
  \BibitemOpen
  \bibfield  {author} {\bibinfo {author} {\bibfnamefont {J.}~\bibnamefont {Sinova}}, \bibinfo {author} {\bibfnamefont {S.}~\bibnamefont {Valenzuela}}, \bibinfo {author} {\bibfnamefont {J.}~\bibnamefont {Wunderlich}}, \bibinfo {author} {\bibfnamefont {C.}~\bibnamefont {Back}}, \ and\ \bibinfo {author} {\bibfnamefont {T.}~\bibnamefont {Jungwirth}},\ }\bibfield  {title} {\enquote {\bibinfo {title} {Spin {H}all effects},}\ }\href {\doibase 10.1103/RevModPhys.87.1213} {\bibfield  {journal} {\bibinfo  {journal} {Rev. Mod. Phys.}\ }\textbf {\bibinfo {volume} {87}},\ \bibinfo {pages} {1213--1260} (\bibinfo {year} {2015})}\BibitemShut {NoStop}%
\bibitem [{\citenamefont {Yamanoi}, \citenamefont {Semizu},\ and\ \citenamefont {Nozaki}(2022)}]{USMR_2}%
  \BibitemOpen
  \bibfield  {author} {\bibinfo {author} {\bibfnamefont {K.}~\bibnamefont {Yamanoi}}, \bibinfo {author} {\bibfnamefont {H.}~\bibnamefont {Semizu}}, \ and\ \bibinfo {author} {\bibfnamefont {Y.}~\bibnamefont {Nozaki}},\ }\bibfield  {title} {\enquote {\bibinfo {title} {Enhancement of room-temperature unidirectional spin {H}all magnetoresistance by using a ferromagnetic metal with a low {C}urie temperature},}\ }\href {\doibase 10.1103/PhysRevB.106.L140401} {\bibfield  {journal} {\bibinfo  {journal} {Phys. Rev. B}\ }\textbf {\bibinfo {volume} {106}},\ \bibinfo {pages} {L140401} (\bibinfo {year} {2022})}\BibitemShut {NoStop}%
\bibitem [{\citenamefont {Ashcroft}(1976)}]{thermoele_1}%
  \BibitemOpen
  \bibfield  {author} {\bibinfo {author} {\bibnamefont {Ashcroft}},\ }\bibfield  {title} {\enquote {\bibinfo {title} {Solid {S}tate {P}hysics},}\ }\href@noop {} {\bibfield  {journal} {\bibinfo  {journal} {Saunders College}\ } (\bibinfo {year} {1976})}\BibitemShut {NoStop}%
\bibitem [{\citenamefont {Goldsmid}(2009)}]{thermoele_2}%
  \BibitemOpen
  \bibfield  {author} {\bibinfo {author} {\bibfnamefont {H.~J.}\ \bibnamefont {Goldsmid}},\ }\bibfield  {title} {\enquote {\bibinfo {title} {Introduction to {T}hermoelectricity},}\ }\href@noop {} {\bibfield  {journal} {\bibinfo  {journal} {Springer-Verlag}\ } (\bibinfo {year} {2009})}\BibitemShut {NoStop}%
\bibitem [{\citenamefont {Bell}(2008)}]{thermoele_3}%
  \BibitemOpen
  \bibfield  {author} {\bibinfo {author} {\bibfnamefont {L.~E.}\ \bibnamefont {Bell}},\ }\bibfield  {title} {\enquote {\bibinfo {title} {Cooling, heating, generating power, recovering waste heat with thermoelectric systems.}}\ }\href {\doibase 10.1126/science.1158899} {\bibfield  {journal} {\bibinfo  {journal} {Science}\ }\textbf {\bibinfo {volume} {321}},\ \bibinfo {pages} {1457--1461} (\bibinfo {year} {2008})}\BibitemShut {NoStop}%
\bibitem [{\citenamefont {Nakai}\ and\ \citenamefont {Nagaosa}(2019)}]{nakainagaosa_PRB}%
  \BibitemOpen
  \bibfield  {author} {\bibinfo {author} {\bibfnamefont {R.}~\bibnamefont {Nakai}}\ and\ \bibinfo {author} {\bibfnamefont {N.}~\bibnamefont {Nagaosa}},\ }\bibfield  {title} {\enquote {\bibinfo {title} {Nonreciprocal thermal and thermoelectric transport of electrons in noncentrosymmetric crystals},}\ }\href {\doibase 10.1103/PhysRevB.99.115201} {\bibfield  {journal} {\bibinfo  {journal} {Phys. Rev. B}\ }\textbf {\bibinfo {volume} {99}},\ \bibinfo {pages} {115201} (\bibinfo {year} {2019})}\BibitemShut {NoStop}%
\bibitem [{\citenamefont {Arisawa}\ \emph {et~al.}(2024)\citenamefont {Arisawa}, \citenamefont {Fujimoto}, \citenamefont {Kikkawa},\ and\ \citenamefont {Saitoh}}]{arisawa_natcomm}%
  \BibitemOpen
  \bibfield  {author} {\bibinfo {author} {\bibfnamefont {H.}~\bibnamefont {Arisawa}}, \bibinfo {author} {\bibfnamefont {Y.}~\bibnamefont {Fujimoto}}, \bibinfo {author} {\bibfnamefont {T.}~\bibnamefont {Kikkawa}}, \ and\ \bibinfo {author} {\bibfnamefont {E.}~\bibnamefont {Saitoh}},\ }\bibfield  {title} {\enquote {\bibinfo {title} {Observation of nonlinear thermoelectric effect in $\rm{MoGe}/ \rm{Y}_3\rm{Fe}_5\rm{O}_{12}$},}\ }\href {\doibase https://doi.org/10.1038/s41467-024-50115-4} {\bibfield  {journal} {\bibinfo  {journal} {Nat. Commun.}\ }\textbf {\bibinfo {volume} {15}},\ \bibinfo {pages} {6912} (\bibinfo {year} {2024})}\BibitemShut {NoStop}%
\bibitem [{\citenamefont {Zhang}\ \emph {et~al.}(2019)\citenamefont {Zhang}, \citenamefont {Lao}, \citenamefont {Skelenar}, \citenamefont {Bingham}, \citenamefont {Batley}, \citenamefont {Watts}, \citenamefont {Nisoli}, \citenamefont {Leighton},\ and\ \citenamefont {Schiffer}}]{Curie_temp}%
  \BibitemOpen
  \bibfield  {author} {\bibinfo {author} {\bibfnamefont {X.}~\bibnamefont {Zhang}}, \bibinfo {author} {\bibfnamefont {Y.}~\bibnamefont {Lao}}, \bibinfo {author} {\bibfnamefont {J.}~\bibnamefont {Skelenar}}, \bibinfo {author} {\bibfnamefont {N.}~\bibnamefont {Bingham}}, \bibinfo {author} {\bibfnamefont {J.}~\bibnamefont {Batley}}, \bibinfo {author} {\bibfnamefont {J.}~\bibnamefont {Watts}}, \bibinfo {author} {\bibfnamefont {C.}~\bibnamefont {Nisoli}}, \bibinfo {author} {\bibfnamefont {C.}~\bibnamefont {Leighton}}, \ and\ \bibinfo {author} {\bibfnamefont {P.}~\bibnamefont {Schiffer}},\ }\bibfield  {title} {\enquote {\bibinfo {title} {Understanding thermal annealing of artificial spin ice},}\ }\href {\doibase https://doi.org/10.1063/1.5126713} {\bibfield  {journal} {\bibinfo  {journal} {APL Mater.}\ }\textbf {\bibinfo {volume} {7}},\ \bibinfo {pages} {111112} (\bibinfo {year} {2019})}\BibitemShut {NoStop}%
\bibitem [{\citenamefont {Meyer}\ \emph {et~al.}(2017)\citenamefont {Meyer}, \citenamefont {Chen}, \citenamefont {Wimmer}, \citenamefont {Althammer}, \citenamefont {Wimmer}, \citenamefont {Schlitz}, \citenamefont {Geprägs}, \citenamefont {Huebl}, \citenamefont {Ködderitzsch}, \citenamefont {Ebert}, \citenamefont {Bauer}, \citenamefont {Gross},\ and\ \citenamefont {Goennenwein}}]{spinNernst_Pt}%
  \BibitemOpen
  \bibfield  {author} {\bibinfo {author} {\bibfnamefont {S.}~\bibnamefont {Meyer}}, \bibinfo {author} {\bibfnamefont {Y.}~\bibnamefont {Chen}}, \bibinfo {author} {\bibfnamefont {S.}~\bibnamefont {Wimmer}}, \bibinfo {author} {\bibfnamefont {M.}~\bibnamefont {Althammer}}, \bibinfo {author} {\bibfnamefont {T.}~\bibnamefont {Wimmer}}, \bibinfo {author} {\bibfnamefont {R.}~\bibnamefont {Schlitz}}, \bibinfo {author} {\bibfnamefont {S.}~\bibnamefont {Geprägs}}, \bibinfo {author} {\bibfnamefont {H.}~\bibnamefont {Huebl}}, \bibinfo {author} {\bibfnamefont {D.}~\bibnamefont {Ködderitzsch}}, \bibinfo {author} {\bibfnamefont {H.}~\bibnamefont {Ebert}}, \bibinfo {author} {\bibfnamefont {G.~E.~W.}\ \bibnamefont {Bauer}}, \bibinfo {author} {\bibfnamefont {R.}~\bibnamefont {Gross}}, \ and\ \bibinfo {author} {\bibfnamefont {S.~T.~B.}\ \bibnamefont {Goennenwein}},\ }\bibfield  {title} {\enquote {\bibinfo {title} {Observation of the spin {N}ernst effect},}\ }\href {\doibase https://doi.org/10.1038/nmat4964} {\bibfield
  {journal} {\bibinfo  {journal} {Nat. Mater.}\ }\textbf {\bibinfo {volume} {16}},\ \bibinfo {pages} {977--981} (\bibinfo {year} {2017})}\BibitemShut {NoStop}%
\bibitem [{\citenamefont {Peng}\ \emph {et~al.}(2017)\citenamefont {Peng}, \citenamefont {Sakuraba}, \citenamefont {Chang}, \citenamefont {Takahashi}, \citenamefont {Mitani},\ and\ \citenamefont {Hayashi}}]{spinNernst_W}%
  \BibitemOpen
  \bibfield  {author} {\bibinfo {author} {\bibfnamefont {S.}~\bibnamefont {Peng}}, \bibinfo {author} {\bibfnamefont {Y.}~\bibnamefont {Sakuraba}}, \bibinfo {author} {\bibfnamefont {Y.}~\bibnamefont {Chang}}, \bibinfo {author} {\bibfnamefont {S.}~\bibnamefont {Takahashi}}, \bibinfo {author} {\bibfnamefont {S.}~\bibnamefont {Mitani}}, \ and\ \bibinfo {author} {\bibfnamefont {M.}~\bibnamefont {Hayashi}},\ }\bibfield  {title} {\enquote {\bibinfo {title} {The spin {N}ernst effect in tungsten},}\ }\href {\doibase 10.1126/sciadv.1701503} {\bibfield  {journal} {\bibinfo  {journal} {Sci. Adv.}\ }\textbf {\bibinfo {volume} {3}},\ \bibinfo {pages} {e1701503} (\bibinfo {year} {2017})}\BibitemShut {NoStop}%
\bibitem [{\citenamefont {Maekawa}\ \emph {et~al.}(2023)\citenamefont {Maekawa}, \citenamefont {Kikkawa}, \citenamefont {Chudo}, \citenamefont {Ieda},\ and\ \citenamefont {Saitoh}}]{maekawa_review_spin_current}%
  \BibitemOpen
  \bibfield  {author} {\bibinfo {author} {\bibfnamefont {S.}~\bibnamefont {Maekawa}}, \bibinfo {author} {\bibfnamefont {T.}~\bibnamefont {Kikkawa}}, \bibinfo {author} {\bibfnamefont {H.}~\bibnamefont {Chudo}}, \bibinfo {author} {\bibfnamefont {J.}~\bibnamefont {Ieda}}, \ and\ \bibinfo {author} {\bibfnamefont {E.}~\bibnamefont {Saitoh}},\ }\bibfield  {title} {\enquote {\bibinfo {title} {Spin and spin current - {F}rom fundamentals to recent progress},}\ }\href {\doibase https://doi.org/10.1063/5.0133335} {\bibfield  {journal} {\bibinfo  {journal} {J. Appl. Phys.}\ }\textbf {\bibinfo {volume} {14}},\ \bibinfo {pages} {020902} (\bibinfo {year} {2023})}\BibitemShut {NoStop}%
\bibitem [{\citenamefont {Cheng}\ \emph {et~al.}(2008)\citenamefont {Cheng}, \citenamefont {Xing}, \citenamefont {Sun},\ and\ \citenamefont {Xie}}]{SNE_NE}%
  \BibitemOpen
  \bibfield  {author} {\bibinfo {author} {\bibfnamefont {S.-g.}\ \bibnamefont {Cheng}}, \bibinfo {author} {\bibfnamefont {Y.}~\bibnamefont {Xing}}, \bibinfo {author} {\bibfnamefont {Q.-f.}\ \bibnamefont {Sun}}, \ and\ \bibinfo {author} {\bibfnamefont {X.~C.}\ \bibnamefont {Xie}},\ }\bibfield  {title} {\enquote {\bibinfo {title} {Spin {N}ernst effect and {N}ernst effect in two-dimensional electron systems},}\ }\href {\doibase 10.1103/PhysRevB.78.045302} {\bibfield  {journal} {\bibinfo  {journal} {Phys. Rev. B}\ }\textbf {\bibinfo {volume} {78}},\ \bibinfo {pages} {045302} (\bibinfo {year} {2008})}\BibitemShut {NoStop}%
\bibitem [{\citenamefont {Liu}\ and\ \citenamefont {Xie}(2010)}]{SNE_theory}%
  \BibitemOpen
  \bibfield  {author} {\bibinfo {author} {\bibfnamefont {X.}~\bibnamefont {Liu}}\ and\ \bibinfo {author} {\bibfnamefont {X.}~\bibnamefont {Xie}},\ }\bibfield  {title} {\enquote {\bibinfo {title} {Spin {N}ernst effect in the absence of a magnetic field},}\ }\href {\doibase https://doi.org/10.1016/j.ssc.2009.12.017} {\bibfield  {journal} {\bibinfo  {journal} {Solid State Communications}\ }\textbf {\bibinfo {volume} {150}},\ \bibinfo {pages} {471--474} (\bibinfo {year} {2010})}\BibitemShut {NoStop}%
\bibitem [{\citenamefont {Kim}\ \emph {et~al.}(2017)\citenamefont {Kim}, \citenamefont {Jeon}, \citenamefont {Choi}, \citenamefont {Lee}, \citenamefont {Surabhi}, \citenamefont {Jeong}, \citenamefont {Lee},\ and\ \citenamefont {Park}}]{hetero_nonlinear}%
  \BibitemOpen
  \bibfield  {author} {\bibinfo {author} {\bibfnamefont {D.}~\bibnamefont {Kim}}, \bibinfo {author} {\bibfnamefont {C.}~\bibnamefont {Jeon}}, \bibinfo {author} {\bibfnamefont {J.}~\bibnamefont {Choi}}, \bibinfo {author} {\bibfnamefont {W.}~\bibnamefont {Lee}, \bibfnamefont {J}}, \bibinfo {author} {\bibfnamefont {S.}~\bibnamefont {Surabhi}}, \bibinfo {author} {\bibfnamefont {J.}~\bibnamefont {Jeong}}, \bibinfo {author} {\bibfnamefont {K.}~\bibnamefont {Lee}}, \ and\ \bibinfo {author} {\bibfnamefont {B.}~\bibnamefont {Park}},\ }\bibfield  {title} {\enquote {\bibinfo {title} {Observation of transverse spin nernst magnetoresistance induced by thermal spin current in ferromagnet/non-magnet bilayers},}\ }\href {\doibase https://doi.org/10.1038/s41467-017-01493-5} {\bibfield  {journal} {\bibinfo  {journal} {Nat. Commun.}\ }\textbf {\bibinfo {volume} {8}},\ \bibinfo {pages} {1400} (\bibinfo {year} {2017})}\BibitemShut {NoStop}%
\bibitem [{\citenamefont {Wu}, \citenamefont {Pearson},\ and\ \citenamefont {Bhattacharya}(2015)}]{lockin_1}%
  \BibitemOpen
  \bibfield  {author} {\bibinfo {author} {\bibfnamefont {S.~M.}\ \bibnamefont {Wu}}, \bibinfo {author} {\bibfnamefont {J.~E.}\ \bibnamefont {Pearson}}, \ and\ \bibinfo {author} {\bibfnamefont {A.}~\bibnamefont {Bhattacharya}},\ }\bibfield  {title} {\enquote {\bibinfo {title} {Paramagnetic spin {S}eebeck effect},}\ }\href {\doibase https://doi.org/10.1103/PhysRevLett.114.186602} {\bibfield  {journal} {\bibinfo  {journal} {Phys. Rev. Lett.}\ }\textbf {\bibinfo {volume} {114}},\ \bibinfo {pages} {186602} (\bibinfo {year} {2015})}\BibitemShut {NoStop}%
\bibitem [{\citenamefont {Vlietstra}, \citenamefont {Shan},\ and\ \citenamefont {Wees}(2014)}]{lockin_3}%
  \BibitemOpen
  \bibfield  {author} {\bibinfo {author} {\bibfnamefont {N.}~\bibnamefont {Vlietstra}}, \bibinfo {author} {\bibfnamefont {J.}~\bibnamefont {Shan}}, \ and\ \bibinfo {author} {\bibfnamefont {B.~J.~V.}\ \bibnamefont {Wees}},\ }\bibfield  {title} {\enquote {\bibinfo {title} {Simultaneous detection of the spin {H}all magnetoresistance and the spin {S}eebeck effect in platinum and tantalum on yttrium iron garnet},}\ }\href {\doibase https://doi.org/10.1103/PhysRevB.90.174436} {\bibfield  {journal} {\bibinfo  {journal} {Phys. Rev. B}\ }\textbf {\bibinfo {volume} {90}},\ \bibinfo {pages} {174436} (\bibinfo {year} {2014})}\BibitemShut {NoStop}%
\bibitem [{\citenamefont {Cornelissen}\ \emph {et~al.}(2017)\citenamefont {Cornelissen}, \citenamefont {Oyanagi}, \citenamefont {Kikkawa}, \citenamefont {Qiu}, \citenamefont {Kuschel}, \citenamefont {Bauer}, \citenamefont {Wees},\ and\ \citenamefont {Saitoh}}]{lockin_4}%
  \BibitemOpen
  \bibfield  {author} {\bibinfo {author} {\bibfnamefont {L.~J.}\ \bibnamefont {Cornelissen}}, \bibinfo {author} {\bibfnamefont {K.}~\bibnamefont {Oyanagi}}, \bibinfo {author} {\bibfnamefont {T.}~\bibnamefont {Kikkawa}}, \bibinfo {author} {\bibfnamefont {W.}~\bibnamefont {Qiu}}, \bibinfo {author} {\bibfnamefont {T.}~\bibnamefont {Kuschel}}, \bibinfo {author} {\bibfnamefont {G.~E.~W.}\ \bibnamefont {Bauer}}, \bibinfo {author} {\bibfnamefont {B.~J.~V.}\ \bibnamefont {Wees}}, \ and\ \bibinfo {author} {\bibfnamefont {E.}~\bibnamefont {Saitoh}},\ }\bibfield  {title} {\enquote {\bibinfo {title} {Nonlocal magnon-polaron transport in yttrium iron garnet},}\ }\href {\doibase https://doi.org/10.1103/PhysRevB.96.104441} {\bibfield  {journal} {\bibinfo  {journal} {Phys. Rev. B}\ }\textbf {\bibinfo {volume} {96}},\ \bibinfo {pages} {104441} (\bibinfo {year} {2017})}\BibitemShut {NoStop}%
\bibitem [{\citenamefont {Avci}\ \emph {et~al.}(2018)\citenamefont {Avci}, \citenamefont {Mendil}, \citenamefont {Beach},\ and\ \citenamefont {Gambardella}}]{USMR_3}%
  \BibitemOpen
  \bibfield  {author} {\bibinfo {author} {\bibfnamefont {C.~O.}\ \bibnamefont {Avci}}, \bibinfo {author} {\bibfnamefont {J.}~\bibnamefont {Mendil}}, \bibinfo {author} {\bibfnamefont {G.~S.~D.}\ \bibnamefont {Beach}}, \ and\ \bibinfo {author} {\bibfnamefont {P.}~\bibnamefont {Gambardella}},\ }\bibfield  {title} {\enquote {\bibinfo {title} {Origins of the unidirectional spin {H}all magnetoresistance in metallic bilayers},}\ }\href {\doibase 10.1103/PhysRevLett.121.087207} {\bibfield  {journal} {\bibinfo  {journal} {Phys. Rev. Lett.}\ }\textbf {\bibinfo {volume} {121}},\ \bibinfo {pages} {087207} (\bibinfo {year} {2018})}\BibitemShut {NoStop}%
\bibitem [{\citenamefont {Zhang}\ and\ \citenamefont {Vignale}(2016)}]{Steven_USMR_theory}%
  \BibitemOpen
  \bibfield  {author} {\bibinfo {author} {\bibfnamefont {S.~S.~L.}\ \bibnamefont {Zhang}}\ and\ \bibinfo {author} {\bibfnamefont {G.}~\bibnamefont {Vignale}},\ }\bibfield  {title} {\enquote {\bibinfo {title} {Theory of unidirectional spin {H}all magnetoresistance in heavy metal/ferromagnetic metal bilayers},}\ }\href {\doibase https://doi.org/10.1103/PhysRevB.94.140411} {\bibfield  {journal} {\bibinfo  {journal} {Phys. Rev. B}\ }\textbf {\bibinfo {volume} {94}},\ \bibinfo {pages} {140411} (\bibinfo {year} {2016})}\BibitemShut {NoStop}%
\bibitem [{\citenamefont {Tumelero}\ \emph {et~al.}(2019)\citenamefont {Tumelero}, \citenamefont {Martins}, \citenamefont {Souza}, \citenamefont {Pace},\ and\ \citenamefont {Pasa}}]{Bi2Se3_Seebeck}%
  \BibitemOpen
  \bibfield  {author} {\bibinfo {author} {\bibfnamefont {M.~A.}\ \bibnamefont {Tumelero}}, \bibinfo {author} {\bibfnamefont {M.~B.}\ \bibnamefont {Martins}}, \bibinfo {author} {\bibfnamefont {P.~B.}\ \bibnamefont {Souza}}, \bibinfo {author} {\bibfnamefont {R.~D.~D.}\ \bibnamefont {Pace}}, \ and\ \bibinfo {author} {\bibfnamefont {A.~A.}\ \bibnamefont {Pasa}},\ }\bibfield  {title} {\enquote {\bibinfo {title} {Effect of electrolyte on the growth of thermoelectric $\rm{Bi_2Se_3}$ thin films},}\ }\href {\doibase https://doi.org/10.1016/j.electacta.2019.01.069} {\bibfield  {journal} {\bibinfo  {journal} {Electrochimica Acta}\ }\textbf {\bibinfo {volume} {300}},\ \bibinfo {pages} {357--362} (\bibinfo {year} {2019})}\BibitemShut {NoStop}%
\bibitem [{\citenamefont {Rakshit}\ \emph {et~al.}(2023)\citenamefont {Rakshit}, \citenamefont {Max}, \citenamefont {Arnab}, \citenamefont {Anthony}, \citenamefont {Xiyue}, \citenamefont {Timothy}, \citenamefont {David}, \citenamefont {Nitin},\ and\ \citenamefont {Daniel}}]{Bi2Se3_spinNernst}%
  \BibitemOpen
  \bibfield  {author} {\bibinfo {author} {\bibfnamefont {J.}~\bibnamefont {Rakshit}}, \bibinfo {author} {\bibfnamefont {S.}~\bibnamefont {Max}}, \bibinfo {author} {\bibfnamefont {B.}~\bibnamefont {Arnab}}, \bibinfo {author} {\bibfnamefont {R.}~\bibnamefont {Anthony}}, \bibinfo {author} {\bibfnamefont {S.}~\bibnamefont {Xiyue}}, \bibinfo {author} {\bibfnamefont {P.}~\bibnamefont {Timothy}}, \bibinfo {author} {\bibfnamefont {A.}~\bibnamefont {David}}, \bibinfo {author} {\bibfnamefont {S.}~\bibnamefont {Nitin}}, \ and\ \bibinfo {author} {\bibfnamefont {C.}~\bibnamefont {Daniel}},\ }\bibfield  {title} {\enquote {\bibinfo {title} {Thermally generated spin current in the topological insulator $\rm{Bi_2Se_3}$},}\ }\href {\doibase https://doi.org/10.1126/sciadv.adi4540} {\bibfield  {journal} {\bibinfo  {journal} {Sci. Adv.}\ }\textbf {\bibinfo {volume} {9}},\ \bibinfo {pages} {eadi4540} (\bibinfo {year} {2023})}\BibitemShut {NoStop}%
\bibitem [{\citenamefont {Pan}\ \emph {et~al.}(2025)\citenamefont {Pan}, \citenamefont {He}, \citenamefont {Feng}, \citenamefont {Li}, \citenamefont {Chen}, \citenamefont {Burkhardt},\ and\ \citenamefont {Felser}}]{Bi88Se12_spinNernst}%
  \BibitemOpen
  \bibfield  {author} {\bibinfo {author} {\bibfnamefont {Y.}~\bibnamefont {Pan}}, \bibinfo {author} {\bibfnamefont {B.}~\bibnamefont {He}}, \bibinfo {author} {\bibfnamefont {X.}~\bibnamefont {Feng}}, \bibinfo {author} {\bibfnamefont {F.}~\bibnamefont {Li}}, \bibinfo {author} {\bibfnamefont {D.}~\bibnamefont {Chen}}, \bibinfo {author} {\bibfnamefont {U.}~\bibnamefont {Burkhardt}}, \ and\ \bibinfo {author} {\bibfnamefont {C.}~\bibnamefont {Felser}},\ }\bibfield  {title} {\enquote {\bibinfo {title} {A magneto-thermoelectric with a high figure of merit in topological insulator $\rm{Bi_{88}Sb_{12}}$},}\ }\href {\doibase https://doi.org/10.1038/s41563-024-02059-9} {\bibfield  {journal} {\bibinfo  {journal} {Nat. Mater.}\ }\textbf {\bibinfo {volume} {24}},\ \bibinfo {pages} {76--82} (\bibinfo {year} {2025})}\BibitemShut {NoStop}%
\bibitem [{\citenamefont {Min}\ \emph {et~al.}(2023)\citenamefont {Min}, \citenamefont {Tan}, \citenamefont {Xie}, \citenamefont {Miao}, \citenamefont {Zhang}, \citenamefont {Lee}, \citenamefont {Gopalan}, \citenamefont {Liu}, \citenamefont {Alem}, \citenamefont {Yan},\ and\ \citenamefont {Mao}}]{strong_nonlinear_Hall}%
  \BibitemOpen
  \bibfield  {author} {\bibinfo {author} {\bibfnamefont {L.}~\bibnamefont {Min}}, \bibinfo {author} {\bibfnamefont {H.}~\bibnamefont {Tan}}, \bibinfo {author} {\bibfnamefont {Z.}~\bibnamefont {Xie}}, \bibinfo {author} {\bibfnamefont {L.}~\bibnamefont {Miao}}, \bibinfo {author} {\bibfnamefont {R.}~\bibnamefont {Zhang}}, \bibinfo {author} {\bibfnamefont {S.}~\bibnamefont {Lee}}, \bibinfo {author} {\bibfnamefont {V.}~\bibnamefont {Gopalan}}, \bibinfo {author} {\bibfnamefont {C.}~\bibnamefont {Liu}}, \bibinfo {author} {\bibfnamefont {N.}~\bibnamefont {Alem}}, \bibinfo {author} {\bibfnamefont {B.}~\bibnamefont {Yan}}, \ and\ \bibinfo {author} {\bibfnamefont {Z.}~\bibnamefont {Mao}},\ }\bibfield  {title} {\enquote {\bibinfo {title} {Strong room-temperature bulk nonlinear {H}all effect in a spin-valley locked {D}irac material},}\ }\href {\doibase https://doi.org/10.1038/s41467-023-35989-0} {\bibfield  {journal} {\bibinfo  {journal} {Nat. Commun.}\ }\textbf {\bibinfo {volume} {14}},\ \bibinfo {pages} {364} (\bibinfo
  {year} {2023})}\BibitemShut {NoStop}%
\bibitem [{\citenamefont {Zhang}\ \emph {et~al.}(2022)\citenamefont {Zhang}, \citenamefont {Lv}, \citenamefont {Wang}, \citenamefont {Zhao}, \citenamefont {Xiong}, \citenamefont {Lv},\ and\ \citenamefont {Zhang}}]{strong_diode_electrical_and_thermal}%
  \BibitemOpen
  \bibfield  {author} {\bibinfo {author} {\bibfnamefont {Y.}~\bibnamefont {Zhang}}, \bibinfo {author} {\bibfnamefont {Q.}~\bibnamefont {Lv}}, \bibinfo {author} {\bibfnamefont {H.}~\bibnamefont {Wang}}, \bibinfo {author} {\bibfnamefont {S.}~\bibnamefont {Zhao}}, \bibinfo {author} {\bibfnamefont {Q.}~\bibnamefont {Xiong}}, \bibinfo {author} {\bibfnamefont {R.}~\bibnamefont {Lv}}, \ and\ \bibinfo {author} {\bibfnamefont {X.}~\bibnamefont {Zhang}},\ }\bibfield  {title} {\enquote {\bibinfo {title} {Simultaneous electrical and thermal rectification in a monolayer lateral heterojunction},}\ }\href {\doibase https://doi.org/10.1126/science.abq0883} {\bibfield  {journal} {\bibinfo  {journal} {Science}\ }\textbf {\bibinfo {volume} {378}},\ \bibinfo {pages} {169--175} (\bibinfo {year} {2022})}\BibitemShut {NoStop}%
\end{thebibliography}%

\end{document}